\documentclass[12 pt]{article}
\usepackage[utf8x]{inputenc}
\usepackage{amsmath}
\usepackage{amsfonts}
\hoffset= - 0.9in         

\newcommand{\p}{\partial}
\newcommand{\f}{\frac}
\newcommand{\s}{\sqrt}

\newcommand{\e}{\epsilon}
\newcommand{\be}{\beta}
\newcommand{\al}{\alpha}

\newcommand{\cf}{\mathcal{F}}

\newcommand{\z}{\zeta}

\newcommand{\mm}{\mathfrak{m}}

\newcommand{\nn}{\nonumber}

\usepackage{hyperref}
\begin{document}
 
 \title{ S-duality as Fourier transform for arbitrary $\e_1,\e_2$}
 \author{{ N.Nemkov}\thanks{{\small
{\it INR, Moscow, Russia} and {\it MIPT, Dolgoprudny, Russia}; nnemkov@gmail.com} }
\date{ }}
\date{\today}
\maketitle
\vspace{-5.0cm}

\begin{center}
 \hfill ITEP/TH-22/13\\
\end{center}

\vspace{3.5cm}
\begin{abstract}
The AGT relations reduce S-duality to the modular transformations
of conformal blocks. It was recently conjectured that for the four-point conformal block the modular transform up to the non-perturbative contributions can be written in form of the ordinary Fourier transform when $\be\equiv-\e_1/\e_2=1$. Here we extend this conjecture to general values of $\e_1,\e_2$. Namely, we argue that for a properly normalized four-point conformal block the S-duality is perturbatively given by the Fourier transform for arbitrary values of the deformation parameters $\e_1,\e_2$.  The conjecture is based on explicit perturbative computations in the first few orders of the string coupling constant $g^2\equiv-\e_1\e_2$ and hypermultiplet masses.   
\end{abstract}
\section{Introduction}
The modular invariance in 2d conformal field theory implies existence of a certain modular transformation of the four-point conformal block. The   AGT correspondence translates the action of this modular transformation to the action of the S-duality on the partition function of the deformed Seiberg-Witten (SW) theory known as the Nekrasov partition function. In a special limit when the deformation parameters $\e_1,\e_2$ are switched off and the Nekrasov partition function is reduced to the partition function of the original SW theory the modular transform is simply the Legendre transform, which is simultaneously the Fourier transform in the limit $\e_1,\e_2\to0$. Hence, we investigate perturbative corrections to the Fourier form of the modular transform in the deformation parameters $\e_1,\e_2$. We demonstrate in several first orders of the perturbation theory that for arbitrary values of the parameters in the theory, i.e. for the general four-point conformal block the modular transform persists to be the Fourier transform. We also propose a natural conjecture that this relation is an exact, valid in all orders of perturbation theory statement.
\subsection*{ Structure of the paper}     
The paper is organized as follows. In section \ref{Sd&Nf} we discuss the S-duality in the deformed SW theory. In section \ref{AGT&MT}, exploiting the AGT correspondence we relate the S-duality to a modular transformation of the conformal block. We see that in the limit restoring the original SW theory this modular transform is the Fourier transform. In sections \ref{Strategy},\ref{PertAnalysis} we describe an approach that allows one to perturbatively calculate the modular kernel. In section \ref{MTasFT} we apply this approach to obtain explicit results.  We find that in every computed order of the perturbation theory the modular transform of the properly normalized conformal block is the Fourier transform. Subtleties concerning the normalization are described in subsequent section \ref{MKnormalization}. Appendices \ref{AppA},\ref{AppB} contain manifest expressions for the pertubatively computed prepotential. Finally, appendices \ref{AppC},\ref{AppD},\ref{AppE} present more detailed and rather technical discussion of the conformal block normalization.      
\section{ S-duality and the Nekrasov partition function\label{Sd&Nf}}
 Seiberg-Witten \cite{SW,SWcont} $\mathcal{N}=2$ SYM $SU(2)$  theory with four fundamental hypermultiplets\footnote{From now on, the name SW theory stands for this particular version.} is an example of the theory enjoying the S-duality \cite{Sdua,Sdua2, Arg}. The S-duality is a transformation relating the strong coupling regime to the weak coupling regime of the same theory. Low energy behaviour of the SW theory is fully encoded in the prepotential function $\cf_{SW}$ which depends on a coordinate $a$ on the vacua moduli space. The action of the $S$-duality on the prepotential function is known and has a simple form of the Legendre transform. Namely, the prepotential of the dual theory $\cf_{SW}^D$ is related to $\cf_{SW} $ by
\begin{equation}
\cf_{SW}^D(b)=2\pi i a_0 b+\cf_{SW}(a_0),\qquad a_0: 2\pi i b+\cf_{SW}'(a_0)=0\label{lt}.
\end{equation}
Another object of interest that has been widely studied recently is a deformation of the original SW theory by the $\Omega$-background \cite{Omega}. The partition function of the deformed theory is the so-called Nekrasov partition function $Z_{Nek}$ \cite{Nek} which depends on two deformation parameters $\e_1,\e_2$. It will be more convenient for us to use instead "the string coupling constant" $g$ and "the $\be$-deformation" $\be$ parameters
\begin{equation}
g^2=-\e_1\e_2,\qquad \be=-\f{\e_1}{\e_2}.
\end{equation}
We keep $\be$ finite  and treat $g$ as a small parameter suitable for a perturbative expansion.

  Prepotential of the undeformed theory $\cf_{SW}$ is obtained from the Nekrasov partition function by taking the limit 
\begin{equation}
\cf_{SW}=\lim_{g\to0}g^2\log {Z_{Nek}}.
\end{equation}
Action of the S-duality should be lifted up to the level of the deformed theory  \cite{GMM}. In this paper, following \cite{GMM}, we propose that the Legendre transform relating $\cf_{SW}$ to its dual \eqref{lt} can be understood as the leading approximation to the Fourier transform relating the Nekrasov partition function $Z_{Nek}$ to its dual. Namely, introduce the prepotential of the deformed theory $F$
\begin{equation}
F =g^2\log{Z_{Nek}}, \qquad \lim_{g\to0}F =\cf_{SW}\label{Fdef},
\end{equation}
 then equation \eqref{lt} appears as the saddle point approximation to  
\begin{equation}
\exp{\left(\f{ F^D(b)}{g^2}\right)}=\int\f{da}{g}\exp{\left(\f{2\pi i a b+F(a)}{g^2}\right)} ,\qquad g\to0\label{ftol}.
\end{equation} 
Analysis presented in this work provides evidences that the Fourier transform is not just the asymptotic form of the $S$-duality in the deformed theory when $g\to0$, but is likely to be the exact relation between the dual theories at the perturbative level.

The problem equivalent to evaluating the S-duality with a glance of non-perturbative contributions was adressed in paper \cite{PT}. Though the explicit formulae for S-duality can be read off from the results obtained in \cite{PT}, formulae there are quite difficult to use in a generic situation.  

What we are going to do, technically, is to investigate not the action of the S-duality on the Nekrasov function itself, but the action of the modular transformation on conformal blocks in 2d conformal field theory. Owing to the AGT relations the S-duality and the modular transform are known to be the same mapping written in the different sets of variables. However, for the conformal block there exists the matrix model representation which is of a great use in the perturbative analysis of the modular transformation. In the next section we establish a link between the S-duality and the modular transformation.
\section{AGT relations and the modular transform\label{AGT&MT}}
Conformal blocks are natural objects in 2d conformal field theory which appear in considerations of multi-point correlation functions \cite{BPZ}. In particular, consider a generic correlator of the four primary fields $V_{\Delta_i}$ with dimensions $\Delta_i$ taken at points $z_i$. Then perform the conformal transformation that maps three of these points in a standard way $(z_1,z_3,z_4)\to (0,1,\infty)$  
\begin{equation}
\big\langle V_{\Delta_1}(z_1)V_{\Delta_2}(z_2)V_{\Delta_3}(z_3)V_{\Delta_4}(z_4)\big\rangle=\prod_{i<j}(z_i-z_j)^{d_{ij}}\big\langle V_{\Delta_1}(0)V_{\Delta_2}(x)V_{\Delta_3}(1)V_{\Delta_4}(\infty)\big\rangle\label{ct},
\end{equation}
here $x$ is the cross-ratio $x=\f{(z_1-z_2)(z_3-z_4)}{(z_1-z_3)(z_2-z_4)}$ and exponents $d_{ij}$ are given by
\begin{align}
&d_{12} = 0,\quad &{d_{13} = \al_1+\al_2+\al_3-\al_4},\quad &{d_{14} = \al_1-\al_2-\al_3+\al_4},\nn\\
&{d_{23} = 0},\quad  &{d_{24} = 2\al_2 },\quad  &{ d_{34} = -\al_1-\al_2+\al_3+\al_4}.
\end{align}
The correlator with the fixed points is commonly represented as a sum over the intermediate dimension $\Delta$
\begin{equation}
\big\langle V_{\Delta_1}(0)V_{\Delta_2}(x)V_{\Delta_3}(1)V_{\Delta_4}(\infty)\big\rangle=\sum_{\Delta} C(\Delta,\Delta_1,\Delta_2)C(\Delta,\Delta_3,\Delta_4)B(\Delta,x,\{\Delta_{1,2,3,4}\})\label{CB},
\end{equation}
where $C$ is structure constant in the three-point correlation function and $B$ is the four-point conformal block.
 
From the AGT relations \cite{AGT,AGTmore,AGTrev} we know that $Z_{Nek}$  coincides with the four-point conformal block $B$ under the proper identification of the gauge theory parameters with the parameters of the conformal theory\footnote{Manifest form of this reparametrization can be found in Appendix \ref{AppD}.}
\begin{equation}
B=Z_{Nek}=\exp{\left(\f{F}{g^2}\right)}\label{B=Nek}.
\end{equation}
   The S-duality in the language of conformal theory becomes a modular transformation of the conformal block. The modular transformation corresponding to the S-duality acts simply as the replacement $x\to1-x$ supplemented with the permutation of the two external dimensions $\Delta_1\leftrightarrow\Delta_3$ \footnote{To simplify the notation we mostly suppress this exchange of the external dimensions in general discussions.}. Thus, slightly abusing the notation and manifestly denoting the $x$-dependence in $F$,  one relates the prepotential to its dual as
\begin{equation}
F^D(a,x)=F(a,1-x).
\end{equation}
 From conformal theory one knows that the action of the  modular transformation on the conformal block can be represented as a linear integral transformation with some modular kernel\footnote{We would like to emphasize that the existence of $x$-independent matrix $M(a,b)$ providing \eqref{mkd} is a non-trivial consequence of special symmetries inherent in the conformal block.} $M(a,b)$
\begin{equation}
\exp{\left(\f{ F(b,1-x)}{g^2}\right)}=\int\f{da}{g}M(a,b)\exp{\left(\f{F(a,x)}{g^2}\right)}\label{mkd}.
\end{equation}
This relation generalizes equation \eqref{ftol}  to arbitrary values of $g$. From equation \eqref{ftol} we know that the asymptotic shape of $M(a,b)$ as $g\to0$ corresponds to the Fourier transform 
\begin{equation}
M(a,b)\underset{g \to 0}{\operatorname{\sim}} \exp{\left(\f{2\pi i a b}{g^2}\right)}\label{amk}.
\end{equation}
The central question we address in this work is how the modular kernel \eqref{amk} modifies when $g\neq0$.
\section{Strategy for modular kernel determination.\label{Strategy}}

 If the prepotential function $F$ or equivalently $Z_{Nek}$ is known, then equation \eqref{mkd} defines the modular kernel $M(a,b)$. Though the Nekrasov partition function is indeed known, there does not seem to be a straightforward way of extracting $M(a,b)$ from \eqref{mkd}. The problem is that currently in a generic case only a representation as a power series expansion in $x$ is available for $Z_{Nek}$. In such representation, the behaviour of $Z_{Nek}$ under the transformation changing $x\to 1-x$ is obscure.

The framework of matrix models offers an approach to the problem. The point is that there exists an integral representation of the conformal block \cite{MMCB,MMCBmore} and a procedure that allows one to compute this integral perturbatively as a power series expansion in the string coupling constant $g$ 
  \begin{equation}
F(a,x)=\sum_{n=0}F_{n}(a,x)g^{2n}\label{pe}
\end{equation}
 with coefficients $F_{n}(a,x)$ being exact and controllable functions of $x$ \cite{GMM}. In the next section it is demonstrated how such representation enables one to construct the modular kernel matrix $M(a,b)$ perturbatively in $g$. More detailed discussion of the matrix model representation for the four-point conformal block is given in Appendix \ref{AppD}.
 
 In fact, due to an involved dependence of the prepotential function $F$ on the hypermultiplet masses $m_f$, one is also forced to treat them perturbatively. If a quantity $t$ parametrizes masses as $m_f=t \mu_f$ then each coefficient function $F_n$ is itself given by a series expansion in $t$
 \begin{equation}
 F_n(a,x)=\sum_{m=0} F_{nm}(a,x)t^{2m}\label{Fnm}.
\end{equation}  
 The dependence of each coefficient $F_{nm}$ on $a$ can be easily restored from the dimensional analysis. The moduli space coordinate $a$, the string coupling constant $g$, and the hypermultiplet masses $m_f$ all having the mass dimension one are the only dimensional parameters in the Nerkasov function. Only positive powers of the coupling constant $g$ and the masses $m_f$ enter prepotential which itself has the mass dimension two. Consequently  
 \begin{equation}
 F_{nm}\propto a^{2(1-n-m)}\label{adep}.
 \end{equation}
 Therefore, expansion \eqref{pe} with $F_n$ given by \eqref{Fnm} can be reorganized into an expansion in the inverse powers of $a$, where $O(a^{-2l})$ stands for terms with $n+m\geq l$. We use the $a$-expansion in presenting particular results, but for general discussions it will be favourable to use the original $g$-expansion \eqref{pe} concealing the expansion in masses.
 
   \section{Perturbative analysis of the modular transform\label{PertAnalysis}}
   Suppose that both the prepotential function $F$ and the modular kernel $M(a,b)$ can be expanded in powers\footnote{The prepotential expansion and consequently the modular kernel expansion contains only even powers of $g$. This can be seen from the manifest expressions for the Nekrasov function as an $x$-series.  }    of $g$
   \begin{eqnarray}
   F(a,x)=\sum_{n=0}g^{2n}F_{n}(a,x),\nn\\
   M(a,b)=\exp\left(\f1{g^2}\sum\limits_{n=0} g^{2n} \mm_n(a,b) \right)\label{expansions}.
   \end{eqnarray}
     The zeroth order coefficient $F_0$ is the prepotential of the undeformed SW theory \eqref{Fdef} $F_0\equiv \cf_{SW}$ and as it follows from \eqref{amk} the zeroth order correction to the modular kernel is \begin{equation}
\mm_0(a,b)      =2\pi i ab.
      \end{equation} 
      
      Substituting expansions \eqref{expansions} into equation \eqref{mkd} one obtains
      \begin{eqnarray}
      \exp{\left(\f1{g^2}\sum\limits_{n=0} g^{2n}F_n(b,1-x)\right)}=\int\f{da}{g}\exp{\left(\f{2\pi i a b+F_0(a,x)}{g^2}\right)}\exp{\left(\f1{g^2}\sum\limits_{n=1} g^{2n}F_n(a,x)\right)}\label{mkexpansion}.
      \end{eqnarray}
   The logarithm of the  l.h.s. of this equation can be straightforwardly expanded in powers of $g$. The integral at the r.h.s. can be evaluated as a power series in $g$ using the saddle point approximation. Saddle point $a_0$ is determined from the equation
  \begin{equation}
   2\pi i b+\p_a F_0(a_0,x)=0\label{saddlepoint}
   \end{equation} 
 and coincides with $a_0$ in equation \eqref{lt}.
 
 Matching expansions of the logarithms of the l.h.s. with the r.h.s. in \eqref{mkexpansion}  then gives the following set of equations 
   \begin{align}
   F_0(b,1-x)=&~2\pi i a_0b+F_0(a_0,x)\label{F0}~,\\
   F_1(b,1-x)=&~\mm_1(a_0,b)+F_1(a_0,x)-\f12\log{\f{F_0''(a_0,x)}{2\pi}}\label{F1},\\
  F_2(b,1-x)=&~F_2(a_0,x)+\mm_2(a_0,b)-\frac{(F_1(a_0,x)+\mm_1(a_0,b))'{}^2}{2 F_0''(a_0,x)}-\frac{(F_1(a_0,x)+\mm_1(a_0,b))''}{2
   F_0''(a_0,x)}+~\nn\\&~\frac{(F_1(a_0,x)+\mm_1(a_0,b))' F_0{}^{(3)}(a_0,x)}{2 F_0''(a_0,x){}^2}-\frac{5
   F_0{}^{(3)}(a_0,x){}^2}{24 F_0''(a_0,x){}^3}+\frac{F_0{}^{(4)}(a_0,x)}{8
   F_0''(a_0,x){}^2}~\label{F2},\\
   F_3(b,1-x)=&~F_3(a_0,x)+\mm_3(a_0,b)+...~\label{F3}
   \end{align}
   where prime denotes derivative w.r.t. $a$.
 
   Note that equation \eqref{F0} states that indeed $F_0$ transforms according to the Legendre transform. Given the first correction to the prepotential $F_1$, the first correction to the modular kernel $\mm_1$ is then straightforwardly determined from equation \eqref{F1} to be
   \begin{equation}
   \mm_1(a_0,b)=F_1(b,1-x)-F_1(a_0,x)+\f12\log{\f{F_0''(a_0,x)}{2\pi}}\label{mk1}.
   \end{equation}
   Having found $\mm_1$, one can then proceed to the second order and so on. Thus, given the first $n$ terms in the prepotential expansion \eqref{pe}, one determines the first $n$ corrections to the modular kernel.
\section{Modular transform as the Fourier transform\label{MTasFT}}
Now we are done with the preliminaries and ready to compute the modular kernel, at least in first few orders of the $g$-expansion. A similar program was carried out in paper \cite{GMM}. The prepotential function $F$ was computed there in the case of $\be=1$ and only two independent hypermultiplet masses up to the order $O(a^{-6})$. Manifest expressions for the coefficient functions
 \begin{equation}
  I_{nm}(a,x)=F_{nm}(\be=1,m_1=m_2,m_3=m_4|a,x)
  \end{equation}
   are presented in Appendix \ref{AppA}. A surprising result was observed. It was found in \cite{GMM} that in this case corrections to the Fourier transform are absent
 \begin{equation}
 M(\be=1,m_1=m_2,m_3=m_4|a,b)=\exp{\left(\f{2\pi i a b+O(a^{-6})}{g^2}\right)}\label{mkr}.
\end{equation}

A natural question is how does \eqref{mkr} generalize to arbitrary values of $m_f$ and $\be$. To figure this out, one needs to compute the prepotential without restrictions on values  of $m_f$ and $\be$ . The matrix model approach is equally suitable in this case. We have extended the results of  \cite{GMM} to arbitrary $m_f$ but fixed $\be=1$. Further generalization to the case of $\be\neq1$ is straightforward. However, in paper \cite{L} by means of a different technique the same amount of corrections as in \cite{GMM} was obtained without constraints on the values of $m_f$ and $\be$ and we exploit these results. Manifest expressions for the prepotential obtained in \cite{L} are listed in Appendix \ref{AppB}. In the case of $\be=1$ they coincide with the prepotential we derived from the matrix models.
   
    Using the prepotential from Appendix \ref{AppB} one can find out how \eqref{mkr} is deformed when the restrictions on $\be$ and $\mu_f$ are lifted. We argue that the Fourier transform shape of the modular kernel stays unchanged in the case of arbitrary values $\be$ and all hypermultiplet masses independent, i.e. for the general four-point conformal block
   \begin{equation}
   M(a,b)=\exp{\left(\f{2\pi i a b+O(a^{-6})}{g^2}\right)}\label{mkramochka}.
\end{equation} 
Moreover, our \emph{conjecture} is that \eqref{mkramochka} is an exact perturbative result, valid in all orders of the $g$-expansion
     \begin{equation}
   \boxed{M(a,b)=\exp{\left(\f{2\pi i a b}{g^2}\right)}\label{conjecture}}
   \end{equation}

Verification of \eqref{mkramochka} is a matter of straightforward substitution of the prepotential from Appendix \ref{AppB} to equations \eqref{F0}-\eqref{F3} with the exception of a subtlety concerning normalization of the conformal block. Normalization issue is discussed in details in the next section.

 One should also notice there was a claim in \cite{GMM} contradicting to \eqref{mkramochka} that when $\be\neq1$ the modular transform is no longer the Fourier transform.  However, as it is explained in the next section the shape of the modular kernel is sensitive to $a$-independent but $x$-dependent renormalizations of the conformal block.  In \cite{GMM} such $a$-independent  contributions to the prepotential were neglected what has led to a deviation from \eqref{mkramochka}. In Appendix \ref{AppE} we revisit the particular case with $\be\neq1$ which was considered in \cite{GMM} and show that restoring the exact $x$-dependence brings this case into agreement with \eqref{mkramochka}. 
\section{Modular kernel and a normalization of the conformal block\label{MKnormalization}}
The prepotential function $F(a,x)$ is usually considered modulo $a$-independent contributions since they do not affect the gauge theory dynamics and such $a$-independent terms were omitted in \cite{GMM},\cite{L}. Hovewer, removal of these terms is equivalent to performing a change of an overall $a$-independent but $x$-dependent normalization of the conformal block, while such renormalizations affect the shape of the modular kernel. Indeed, suppose that equation \eqref{mkd} is satisfied by some conformal block $B(a,x)$ and by some modular kernel $\mathcal{M}(a,b)$
\begin{equation}
B(b,1-x)=\int da\, \mathcal{M}(a,b)\, B(a,x)\label{NS}.
\end{equation}  
Then a renormalization of this conformal block  $\widetilde{B}(a,x)=N(x)B(a,x)$ by some function $N(x)$ breaks \eqref{NS} with the original kernel $\mathcal{M}(a,b)$ unless $N(x)=N(1-x)$. Moreover, for a generic $N(x)$ there does not exist any $x$-independent kernel providing modular transformation. However, for some functions $N(x)$ it is possible to redefine the kernel in order to retain \eqref{NS}
\begin{equation}
\widetilde{B}(b,1-x)=\int da\,\widetilde{\mathcal{M}}(a,b)\widetilde{B}(a,x).
\end{equation}
It turns out that removal of some $a$-independent terms in the prepotential function is the renormalization of such kind, i.e. there still exists a linear $x$-independent transformation relating truncated prepotential at the points $x$ and $1-x$ but with the kernel different from that for the full prepotential. 

In Appendix \ref{AppC} we demonstrate that it is possible to renormalize the truncated prepotential from Appendix \ref{AppB} so that the modular kernel is the Fourier kernel in all computed orders. 

In Appendix \ref{AppD} by means of the matrix model representation, the exact normalization originally present in the conformal block is computed. Pleasantly, it appears to be the same normalization in which the modular transform is the Fourier transform. Note that there may be an ambiguity of whether attributing factor $\prod_{i<j}(z_i-z_j)^{d_{ij}}$ from equation \eqref{ct} to the conformal block or not. However, this factor is symmetric under the simultaneous exchange of $z_1\leftrightarrow z_3$ (which generates $x\to 1-x$) and $\Delta_1\leftrightarrow\Delta_3$. Therefore this normalization uncertainty is irrelevant for the shape of the modular transform. Thus, the conformal block normalized in the standard way transforms according to the Fourier transform in all computed orders \eqref{mkramochka}.

\section{Conclusion\label{Conclusion}}
In this paper we performed a perturbative analysis of the S-duality in the deformed SW theory. This was done by relating the S-duality transformation to the modular transformation of conformal blocks in 2d conformal field theory and using the matrix model representation of the conformal block together with the results obtained in \cite{L}. We found that in the first several orders of the perturbation theory the modular transform of the general four-point conformal block normalized in the standard way is given by the ordinary Fourier transform. We have conjectured that the Fourier kernel is the exact perturbative form of the modular kernel.

 In paper  \cite{PT} exact answer for the modular kernel including non-perturbative corrections was presented. Unfortunately, the explicit expressions obtained there seem to be rather difficult to use. We are currently unable to check these formulae against our conjecture for the perturbative shape of the modular kernel in the general case. It is not even clear how to separate perturbative from the non-perturbative parts in the answer from \cite{PT}.   
 
 {\bf Note added.} When this paper was completed, we became aware of paper \cite{to appear}, where the authors also address
the problem of constructing the modular transformation including the non-perturbative corrections. They explicitly solve the problem in the case of
the unit central charge only and seem to confirm that, in this case, the perturbative corrections are absent.
Putting together all these results which are complementary to each other might help to gain more conceptual understanding of the observed phenomena
and is the matter of the future work.
\section*{Acknowledgements}
The author is grateful to D. Galakhov, A. Mironov, and M.Morozov for numerous fruitful discussions. This work is partly supported by the RFBR grant 13-02-00457.
    \appendix   
\section{Prepotential from matrix models\label{AppA}}
Here several first terms computed in \cite{GMM} by means of the matrix model description are presented. Corrections are parametrized according to
\begin{equation}
F(\beta=1,m_1=m_2=t\mu_1,m_3=m_4=t\mu_3|a,x)=\sum_{n,m=0}I_{nm}(a,x)g^{2n}t^{2m}.
\end{equation}
Several first coefficients $I_{nm}$ are given by
 \begin{align}
I_{00}=&~-\frac{\pi a^2 K(1-x)}{K(x)}\label{F00}~,\nn\\
I_{01}=&~2(\mu_1^2+\mu_3^2)\log a~\nn,\\
I_{02}=&~-\frac{2(\mu_1^4+\mu_3^4)}{3\pi^2 a^2}K(x)\left((x-2)K(x)+3E(x)\right)-\frac{4\mu_1^2\mu_3^2}{\pi^2a^2}K(x)\left((x-1)K(x)+E(x)\right)~\nn,\\
I_{10}=&~-\frac{1}{2}\log a~\nn,\\
I_{11}=&~\frac{2(\mu_1^2+\mu_3^2)}{3\pi^2 a^2}K(x)\left((x-2)K(x)+3E(x)\right)~\nn,\\
I_{12}=&~\frac{4(\mu_1^4+\mu_3^4)}{3\pi^2 a^4}K^2(x)\left[(x^2-3x+3)K^2(x)+4(x-2)K(x)E(x)+6E^2(x)\right]+\nn~\\&~
\frac{8\mu_1^2 \mu_3^2}{3\pi^4 a^4}\left[(3x^2-7x+4)K^2(x)+2(4x-5)K(x)E(x)+6E^2(x)\right]~\nn,\\
I_{20}=&~-\frac{K(x)}{8\pi^2 a^2}\left((x-2)K(x)+3E(x)\right)~\nn,\\
I_{21}=&~-\frac{\mu_1^2+\mu_3^2}{60\pi^4 a^4}K^2(x)\left((48 x^2-143x+143)K^2(x)+190(x-2)K(x)E(x)+285 E^2(x)\right)~\nn,\\
I_{22}=&~- \frac{(\mu_1^2+\mu_3^2)^2}{90\pi^6 a^6} K^3(x)\left(-1646 K^3(x)+4350 E^3(x)+2469 x K^3(x)-8700 K(x) E^2(x)+\nn\right.~ \nn\\ &~\left.6476 K^2(x) E(x) -1783 K^3(x)^3 x^2+480 x^3 K^3(x)+4350 K(x) E^2(x) x+2126 K^2(x) x^2 E(x)-~\right.\nn\\ &~\left.  6476 K^2(x) x E(x)\right)~ .
\end{align}
Here $K(x),E(x)$ are the complete elliptic integrals.
\section{Prepotential from the work \cite{L}\label{AppB}}
The prepotential computed in \cite{L} is presented. Corrections are parametrized according to 
\begin{equation}
F=\sum_{n,m=0}(\e_1+\e_2)^{2n}(\e_1\e_2)^{2m}\cf^{(n,m)}\equiv\sum_{n,m=0}g^{2n+2m}(\s{\be}-\f1{\s{\be}})^{2n}(-1)^m \cf^{(n,m)}.
\end{equation}
\begin{align}
 \cf^{(0,0)}=&~ 2 R \log\frac{a}{\Lambda}-\frac{R^2E_2}{6a^2}+\frac{T_1\theta_4^4-T_2\theta_2^4}{a^2}
-\frac{R^3(5E_2^2+E_4)}{180\,a^4}-\frac{NE_4}{5\,a^4}\notag\\
&~+\frac{RT_1\theta_4^4(2E_2+2\theta_2^4+\theta_4^4)}{6\,a^4}-
\frac{RT_2\theta_2^4(2E_2-2\theta_4^4-\theta_2^4)}{6\,a^4}
+\cdots~, \nn\\
\cf^{(1,0)}=&~-\frac{1}{2}\log \frac{a}{\Lambda}+\frac{R E_2}{12\,a^2}
+\frac{R^2(E_2^2+E_4)}{48\,a^4}\notag\\
&~-\frac{T_1\theta_4^4(E_2+4\theta_2^4+2\theta_4^4)}{12\,a^4}+
\frac{T_2\theta_2^4(E_2-4\theta_4^4-2\theta_2^4)}{12\,a^4}+\cdots~, \nn\\
\cf^{(0,1)}=&~\frac{1}{2}\log\frac{a}{\Lambda}-\frac{RE_2}{6\,a^2}
-\frac{R^2(2E_2^2+E_4)}{36\,a^4}\notag\\
&~+\frac{T_1\theta_4^4(2E_2+2\theta_2^4+\theta_4^4)}{6\,a^4}-
\frac{T_2\theta_2^4(2E_2-2\theta_4^4-\theta_2^4)}{6\,a^4}+\cdots~, \nn\\
\cf^{(2,0)}=&~-\frac{E_2}{96a^2}-\frac{R(5E_2^2+9E_4)}{960\,a^4}+\cdots~, \nn\\
\cf^{(1,1)}=&~\frac{E_2}{24a^2}+\frac{R(10E_2^2+11E_4)}{360\,a^4}+\cdots~, \nn\\
\cf^{(0,2)}=&~-\frac{E_2}{32a^2}-\frac{R(95E_2^2+49E_4)}{2880\,a^4}+\cdots~, \nn\\
\cf^{(3,0)}=&~\frac{5E_2^2+13E_4}{11520\,a^4}\cdots~,\qquad\qquad~ \cf^{(2,1)}=-\frac{10E_2^2+17E_4}{2880\,a^4}\cdots~, \nn\\
\cf^{(1,2)}=&~\frac{95E_2^2+94E_4}{11520\,a^4}\cdots~,\qquad\qquad \cf^{(0,3)}=-\frac{2E_2^2+E_4}{384\,a^4}\cdots~. 
\end{align}
Here the quantities $R,T_1,T_2,N$ parametrise the dependence on the hypermultiplets masses
\begin{equation}
 \label{invdef}
 \begin{aligned}
 R & = \frac 12 \sum_f m_f^2
~,\\
  T_1 & = \frac{1}{12} \sum_{f<f'} m_f^2 m_{f'}^2 - \frac{1}{24} \sum_f m_f^4
~,\\
 T_2 & = -\frac{1}{24} \sum_{f<f'} m_f^2 m_{f'}^2 + \frac{1}{48} \sum_f m_f^4 
-\frac 12 \prod_f m_f
~,\\
N & = \frac{3}{16} \sum_{f<f'<f''} m_f^2 m_{f'}^2 m_{f''}^2 - \frac{1}{96} 
 \sum_{f\not= f'} m_f^2 m_{f'}^4 + \frac{1}{96} \sum_f m_f^6~.
\end{aligned} 
\end{equation} 
The special functions entering this expressions are related to the complete elliptic integrals $K(x),E(x)$ as follows
\begin{align}
E_2(q)=\f{4K(x)}{\pi^2}((x-2)K(x)+3E(x)),\quad
E_4(q)=\f{16(x^2-x+1)K^4(x)}{\pi^4},\nn\\
\theta_2^4(q)=\f{4xK(x)^2}{\pi^2},\quad
\theta_4^4(q)=\f{4(1-x)K(x)^2}{\pi^2},
\end{align}
where $q$ is related to $x$ by
\begin{equation}
x=\f{\theta_2^4(q)}{\theta_3^4(q)}
\end{equation}
and $\theta_{2,3,4}$ are the Jacobi theta functions.

\section{Normalization providing Fourier transform\label{AppC}}
In this appendix we illustrate in detail how does the truncation of $x$-independent terms in the prepotential change the modular kernel. We also demonstrate that it is possible to normalize the truncated prepotental from Appendix \ref{AppB} in order for the modular kernel to be the Fourier kernel in all computed orders.
 
In the perturbative expansion $a$-independent function may appear only in $F_{10}$ or $F_{01}$ \eqref{Fnm} corrections, since they are the only ones allowed for a trivial dependence on $a$ (see eq. \eqref{adep}). We accurately examine $F_{10}$ correction and the corresponding contribution to the modular kernel \footnote{We define $\mm_n(a,b)=\sum\limits_{m=0}\mm_{nm}(a,b)t^{2m}$ in analogy with \eqref{Fnm} } $\mm_{10}$. The other correction $F_{01}$ can be dealt with in the very same way. 

To determine $\mm_{10}$ from equation \eqref{mk1} one needs $F_{00}$, $F_{10}$ and $a_0$. From Appendix \ref{AppA}
\begin{equation}
F_{00}=I_{00}=-\f{\pi a^2 K(1-x)}{K(x)}.
\end{equation} 
The saddle point $a_0$ satisfying equation \eqref{saddlepoint} is then
\begin{equation}
a_0=ib\f{K(x)}{K(1-x)}\label{a0}.
\end{equation}
Introduce the notation
 \begin{equation}
 F_{10}(a,x)=\widetilde{F}_{10}(a,x)+f_{10}(x)\label{IN}.
\end{equation} 
  Here $\widetilde{F}_{10}$ is $F_{10}$ with $a$-independent part omitted and $f_{10}$ is a function of $x$ solely. Then from Appendix \ref{AppB} 
\begin{equation}
\widetilde{F}_{10}=\left(\s{\be}-\f1{\s{\be}}\right)^2\cf^{(1,0)}|_{m_f=0}-\cf^{(0,1)}|_{m_f=0} =-\f{\be^2-\be+1}{2\be}\log{a}. 
\end{equation} 
 Substituting this to \eqref{mk1} one yields 
\begin{eqnarray}
 \mm_{10}(a_0,b)=F_{10}(b,1-x)-F_{10}(a_0,x)+\f12\log{\f{F_0''(a_0,x)}{2\pi}}&=&\nn\\
-\f{(\be-\be+1)}{2\be}\log{\f{b}{a_0}}+\f12\log{\f{K(1-x)}{K(x)}}+f_{10}(1-x)-f_{10}(x)&=&\nn\\ -\f{(\be-1)^2}{2\be}\log{\f{b}{a_0}}+f_{10}(1-x)-f_{10}(x).\label{m10particular}
\end{eqnarray}
When passing from the second to the third line we translated the explicit dependence on $x$ in the term $\log{K(1-x)/K(x)}$ to the dependence on $a_0(x)$ via relation \footnote{We have skipped the imaginary unit from \eqref{saddlepoint} since it would only contribute as a numerical factor to $M(a,b)$. Hereafter, we always do this.}\eqref{a0}
\begin{equation}
\log{\f{K(1-x)}{K(x)}}=\log{\f{b}{a_0}}.
\end{equation}
The same elimination of $x$ must be possible for the combination\footnote{This is not true for an arbitrary function $f_{10}$. The fact that the elimination is possible is a manifestation of the $x$-independence of the modular kernel $M(a,b)$ in \eqref{mkd}.}
\begin{equation}
f_{10}(1-x)-f_{10}(x)\equiv \delta\mm_{10}(a_0,b).
\end{equation} 
Thus we see that the presence of function $f_{10}$ affects the $\mm_{10}$ correction to the modular kernel. Moreover, since $\mm_{1}$ enters the equations for the higher orders \eqref{F2},\eqref{F3},... all orders of the modular kernel are sensitive to $f_{10}$. The same reasoning covers $F_{01}$ and $\mm_{01}$ corrections. 
  
  There are no such ambiguities in the higher orders, and other corrections listed in Appendix \ref{AppB} are exact. The key observation is that equations \eqref{F0}-\eqref{F3} are satisfied by $F$ from Appendix \ref{AppB} and $\mm_0=2\pi iab,\mm_1=\mm_2=\mm_3...=0$ modulo $a$-independent functions. In other words, the higher order corrections $F_{nm},n+m\geq2$ are related to each other just as if the modular transform was the Fourier transform. However, a "mismatch" in the lowest orders leading to non-vanishing $\mm_{10}$ and $\mm_{01}$ affects the higher orders of the modular kernel and in the case of $\be\neq1$ ruins the Fourier shape of the modular kernel \eqref{mkramochka} as it was found in \cite{GMM} and was illustrated above.
     
  This "mismatch" in the prepotential from Appendix \ref{AppB} can be cured by an appropriate choice of the functions $f_{10},f_{01}$.  In fact, such functions are originally presented in the conformal block normalized in the standard way, for instance as in \eqref{be}. This statement is highlighted in Appendix \ref{AppD}. Below we demonstrate that it is indeed possible to choose $f_{10}$ so that $\mm_{10}=0$.
  
  Equation \eqref{m10particular} asserts that in order for $\mm_{10}$ to vanish, $f_{10}$ must satisfy 
  \begin{equation}
  f_{10}(1-x)-f_{10}(x)= \f{(\be-1)^2}{2\be}\log{\f{b}{a_0}}= \f{(\be-1)^2}{2\be}\log{\f{K(1-x)}{K(x)}}\label{f10}.
  \end{equation}
Since the r.h.s. of this equation is antisymmetric w.r.t. the change $x\to1-x$, there is a solution to \eqref{f10}
\begin{equation}
f_{10}(x)=-\f{(\be-1)^2}{4\be}\log{\f{K(1-x)}{K(x)}}+g_{10}(x),
\end{equation}
where $g_{10}$ is an arbitrary function\footnote{Appearing of such arbitrariness is in no way a surprise since it is obvious that if $F(a,x)$ fulfils  equation \eqref{mkd} then $F(a,x)+g^2g_{10}(x)$ also does.} satisfying $g_{10}(x)=g_{10}(1-x)$. 

The very same story happens with $\mm_{01}$ correction: there is a choice of $f_{01}$ (differing from $f_{10}$ by a numerical constant) providing $\mm_{01}=0$. Particular shapes of functions $f_{10},f_{01}$ for a certain normalization of the conformal block are presented in Appendix \ref{AppD}.

 Hereby we see that there is a normalization of the conformal block in which $\mm_{10}=\mm_{01}=0$ and as we've claimed earlier this leads to \eqref{mkramochka} without any additional ambiguities. 
 
\section{Matrix model representation of the conformal block\label{AppD}}
In this appendix we give a brief description of the matrix-model representation of the conformal block, sketch the procedure of computing prepotential $F$ perturbatively in powers of $g$, and comment on why this procedure allows to accurately account for $a$-independent normalizations.

$\be$-ensemble partition function is the matrix model\footnote{The $\be$-ensemble is not literally a matrix model for $\be\neq1$ but rather an eigenvalue integral. However, there is no essential difference for our considerations and we will keep referring to the $\be$-ensemble as to a matrix model.} representation of the four-point conformal block. It is defined by  
  \begin{equation}
 Z_{\be}=\prod\limits_{a<b}(q_a-q_b)^{\f{2\al_a\al_b}{g^2}}\oint_{\gamma_i}dz_i \prod_{i<j}z_{ij}^{2\be}\prod_i\prod_{a} (z_i-q_a)^{\f{2\s{\be}\al_a}{g}}\label{be}.
 \end{equation}
where $q_a=\{0,x,1\}$, among contours $\gamma_i$ there are $N_1$ segments $[0,x]$ and $N_2$ segments $[0,1]$, with $N_1,N_2$ equal to
\begin{eqnarray}
N_1=\f1{\s{\be}}\left(\f{\al-\al_1-\al_2}{g}\right),\qquad
N_2=\f1{\s{\be}}\left(\s{\be}-\f1{\s{\be}}-\f{\al+\al_3+\al_4}{g}\right).
\end{eqnarray}
 
The $\be$-ensemble partition function provides an integral representation of the conformal block $B$ defined in $\eqref{CB}$ $B=Z_\be$ \cite{MMCB,MMCBmore}. Thereby, together with \eqref{B=Nek} this relates the three different objects 
\begin{equation}
B=Z_{Nek}=Z_\be.
\end{equation}

Relations between the parameters entering $\be$-ensemble and the conformal/gauge theory variables can be assigned through \cite{AGTrev}
\begin{eqnarray}
\Delta(\al)=\f{\al(\e-\al)}{\e_1\e_2}\nn,\\
\e_1\e_2=-g^2,\quad \e_1/\e_2=-\be,\quad
\al=a+\e/2,\quad x=e^{2\pi i \tau_0}\nn,\\
m_1=\al_1+\al_2,\quad m_2=\al_1-\al_2+\e,\quad m_3=\al_3+\al_4,\nn\\\quad m_4=\al_3-\al_4+\e.
\end{eqnarray}
where $\e=\e_1+\e_2$, $m_f$ are the hypermultiplets masses, $a$ is the vacua moduli space coordinate and $\tau_0$ is the bare coupling constant of the gauge theory.

The logarithm of the $\be$-ensemble partition function  \footnote{We use the same notation $F$ for the logarithms of $Z_\be$ and $Z_{Nek}$ since the AGT relations implies their coincidence.}\eqref{be} $F=g^2\log{Z_\be}$ satisfies the Seiberg-Witten equations
\begin{eqnarray}
a=\f1{2\pi i}\oint_A\Omega_{\e_1,\e_2},\quad
\p_a F=\oint_B\Omega_{\e_1,\e_2}\label{SW}.
\end{eqnarray}
The subscripts $\e_1,\e_2$ emphasize the validity of these equations in the deformed theory. The role of Seiberg-Witten differential is played by the one-point resolvent $r_1$ of the $\be$-ensemble
\begin{eqnarray}
\Omega_{\e_1,\e_2}=r_1(\z)d\z, \quad
r_1(\z)=\left\langle \sum_{i}\f1{\z-z_i}\right\rangle_{Z_\be},
\end{eqnarray}
where $\left\langle\cdots\right\rangle_{Z_\be}$ denotes the average with respect to the measure in \eqref{be}. 

The one-point resolvent $r_1$ in its turn can be restored from the so-called loop equations \cite{loop equations 1,loop equations 2}. Namely, if one also introduces the multi-point resolvents $r_n$
\begin{equation}
r_n(\z_1,...,\z_n)=\left\langle \sum_{i_1}\f1{\z_1-z_{i_1}}\cdots \sum_{i_n}\f1{\z_n-z_{i_n}}\right\rangle_{Z_\be},
\end{equation}    
then the loop equations for the $\be$-ensemble are
\begin{align}
\be r_{n+1}(\z,\z,x_1,...,x_{n-1})+(\be-1)\p_\z r_n(\z,x_1,...,x_{n-1})+\nn\\\sum_{a}\f{2\s{\be} \al_a}{g}\f{r_n(\z,x_1,...,x_{n-1})-r_n(q_a,x_1,...,x_{n-1})}{\z-q_a}+\nn\\\sum_i\p_{x_i}\f{r_{n-1}(x_1,...,x_{n-1})-r_{n-1}(x_1,...,x_n)|_{x_i=q_a}}{x_i-\z}=0\label{led}.
\end{align} 
These equations are essentially the Virasoro constraints. For $n=1$ the first term in the second line in equation \eqref{led} contains the one-point resolvent $r_1$ while the second term can be written as
\begin{eqnarray}
\sum_{a}\f{2\s{\be} \al_a}{g}\f{r_1(q_a)}{\z-q_a} = \sum_a\f{\p_{q_a}F(a,x)}{\z-q_a}=\f{(x^2-x)}{z(z-x)(z-1)}\p_x F(a,x)+...
\end{eqnarray}
Here, the derivatives w.r.t. $q_a$ are meant to be taken in \eqref{be} before setting $q_a=\{0,x,1\}$. Dots stay for the term that appears from the accurate restoring of the $q_a$-dependence. This term is not important for this qualitative discussion, though is important for exact calculations.   The point to emphasize here is that the equation for the one-point resolvent includes the derivative of $F$ w.r.t. $x$. Therefore, the one-point resolvent and consequently the SW equations \eqref{SW} are sensitive to $a$-independent contributions in $F$. This is the technical reason why the $\be$-ensemble representation allows one to find $F$ precisely accounting for such $a$-independent normalization. 
Furthermore, the loop equations \eqref{led} rewritten in terms of the connected resolvents enable to construct $r_1$ as a power series expansion in $g$ (in order to find $n$-th correction to $r_1$ one only needs to solve the first $n$ loop equations for the connected resolvents). Thereby, by means of the matrix model description one can compute the prepotential perturbatively in $g$ accurately accounting for the $a$-independent terms.

Particular calculations show that the functions $f_{10}$ and $f_{01}$ \eqref{IN} corresponding to the normalization chosen in \eqref{be} are
\begin{eqnarray}
f_{10}(x)=-\frac{(\beta-1 )^2 }{2 \beta } 
   \log{K(x)} -\frac{(\beta-1 )^2 }{4 \beta } \log{(x^2-x)}\nn,\\
   f_{01}(x)=-\left(m_1^2+m_2^2+m_3^2+m_4^2\right) \log{K(x)}-\frac{1}{2}
   \left(m_1^2+m_2^2\right) \log{x}-\nn\\\frac{1}{4} \left(m_1^2+m_2^2+m_3^2+m_4^2-2 m_2
   m_1+2 m_3 m_4\right) \log{(x-1)}\label{f10,f01}.
\end{eqnarray}
   One can simply check that these functions accurately provide $\mm_{10}=\mm_{01}=0$. Note that in each of these functions only the first term affects the modular kernel while the remainders are symmetric with respect to the exchange $x\leftrightarrow1-x$ and\footnote{This is the permutation of the external dimensions that supplements the change $x\to1-x$ in the modular transformation which we previously did not mention explicitly.} $\al_1\leftrightarrow\al_3$. Therefore, the modular transformation for the conformal block \eqref{be} is the Fourier transform in all computed orders.
\section{Modular kernel for the truncated prepotential\label{AppE}}
   In \cite{GMM} the conformal block with \footnote{Note that the notation in the present paper slightly differs from \cite{GMM}. Namely, $m_f|_{\mbox{here}}=\mu_f|_{\mbox{there}}+\e/2$ .} $m_1=m_3=0, m_2=m_4=\e$ was chosen as an illustration to the sample of $\be\neq1$. From Appendix \ref{AppB} we see that in this case the first two corrections to the truncated prepotential $\widetilde{F}$ are
   \begin{eqnarray}
   \widetilde{F}_1&=& \f{1}{2}(3\be -7 +\f{3}{\be})\log{a},\\
   \widetilde{F}_2&=&-\f{1}{8\pi^2a^2}\left(3\be-7+3\f1{\be}\right)\left( \be-3+\f1{\be}\right)K(x)((x-2)K(x)+3E(x))\label{tF2}.
       \end{eqnarray}
       in agreement with what was found in \cite{GMM}.
  The corresponding contributions to the modular kernel $\widetilde{\mm}_1,\widetilde{\mm}_2$ are determined from \eqref{F1},\eqref{F2} and are equal to
  \begin{eqnarray}
  \widetilde{\mm}_1(a,b)=\f{3(\be-1)^2}{2\be}\log{\f{b}{a}}\label{m1be},\\
   \widetilde{\mm}_2(a,b)=-\f{3i}{16\pi ab}\f{(\be-3)(3\be-1)(\be-1)^2}{\be^2}
  \end{eqnarray}
 again in agreement with \cite{GMM}. 
 
 Now, restoring of the $a$-independent terms according to \eqref{f10,f01} gives
  \begin{eqnarray}
  F_1&=&\widetilde{F}_1-\f{3(\be-1)^2}{2\be}\log{K_x}+...,\\
  F_2&=&\widetilde{F}_2.
  \end{eqnarray}
  Here dots stay for the terms symmetric under the change $x\to1-x, \Delta_1\leftrightarrow\Delta_3$ and thus irrelevant for the modular transformation. It is easily seen that the effect of the additional $a$-independent function is to exactly cancel \eqref{m1be} so that for the full correction $F_1$ the corresponding correction to the modular kernel vanishes $\mm_1=0$. Furthermore, equation \eqref{F2} with $\mm_1=0$ and $F_2=\widetilde{F}_2$ from \eqref{tF2} gives $\mm_2=0$. Thereby, we have explicitly demonstrated that restoring the $a$-independent terms in the case of $\be\neq1$ changes the shape of the modular kernel found in \cite{GMM} to the Fourier kernel.

  
\end{document}